\documentclass[reqno,10pt,centertags]{amsart}
\usepackage{amssymb,upref,graphicx,lastpage}
\global\arraycolsep=2pt
\newcommand{\be}{\begin{equation}}
\newcommand{\ee}{\end{equation}}
\newcommand{\bea}{\begin{eqnarray}}
\newcommand{\eea}{\end{eqnarray}}

\allowdisplaybreaks \numberwithin{equation}{section}

\usepackage{graphicx,lastpage}
\makeatletter
\def\@maketitle@hook
  {\hbox to \textwidth
    {\valign{\vss##\vss\cr
      \noalign{\hfil}
      \hsize=0.7\hsize
      \vbox{\parindent0pt \raggedleft

         \medskip
         \sffamily
         Transactions of The\\ Royal Norwegian Society\\ of Sciences and
         Letters,\\ 
         (\textit{Trans.\ R.\ Norw.\  Soc.\  Sci.\ Lett.} 20xx(x) x--x)\par}\cr}}
  \vskip 42pt \relax}
\makeatother

\begin{document}

\title[]{Casimir torque in weak coupling}

\author[Milton]{Kimball A. Milton}

\email{milton@nhn.ou.edu}
\urladdr{nhn.nhn.ou.edu/\%7Emilton}

\author[Parashar]{Prachi Parashar}
\email{prachi@nhn.ou.edu}
\address{Homer L. Dodge Department of Physics and Astronomy, University of Oklahoma,
Norman, OK 73019 USA}

\author[Long]{William Long}

\dedicatory{Dedicated  to Johan H\o ye}
\thanks{This work was supported by the US National Science Foundation and
the Julian Schwinger Foundation. We thank Elom Abalo, Nick Pellatz,
 and Nathan Yu for collaborative assistance.}
\date{\today}

\begin{abstract}
In this paper, dedicated to Johan H\o ye on the occasion of his 70th birthday,
we examine manifestations of Casimir torque in the weak-coupling approximation,
which allows exact calculations so that comparison with the universally
applicable, but generally uncontrolled, proximity force approximation may be
made.  In particular, we examine Casimir energies between planar objects
characterized by $\delta$-function potentials, and consider the torque
that arises when angles between the objects are changed.
The results agree very well with the proximity force approximation
when the separation distance between the objects is small compared
with their sizes. In the opposite limit, where the size of one object
is comparable to the separation distance, the shape dependence starts
becoming irrelevant.  These calculations
are illustrative of what to expect for the torques between, for example,
conducting planar objects, which eventually should be amenable to both
improved theoretical calculation and experimental verification.
\end{abstract}

\maketitle

\section{Introduction}
The forces due to quantum field fluctuations between parallel 
planar surfaces have
been studied theoretically for many years, first for perfect conductors by 
Casimir \cite{Casimir:1948dh}, and for dielectrics by Lifshitz et al.\
\cite{Lifshitz:1956zz,dzyaloshinskii}.  The subject has reached a mature stage,
with many precision experimental investigations, and a variety of applications;
for recent reviews see Ref.~\cite{Bordag:2009zz,dalvitbook}.

There have been a number of previous considerations of Casimir torque.
For example, the torque between corrugated cylinders was proposed in
Ref.~\cite{Lombardo:2008zza}, and computed perturbatively in 
Ref.~\cite{CaveroPelaez:2008tk}.  A very interesting calculation
of torque between birefringent plates was made a number of years ago by
Barash \cite{barash}, and updated more recently in Ref.~\cite{barash2}.
To our knowledge, none of these effects have been observed, although
the related lateral force between corrugated surfaces has been seen in
experiments \cite{lateral1,lateral2}.
 
In this paper we will consider the torque due to fluctuations in
 a scalar field, where the corresponding
Green's function is defined by the differential equation
\be
(-\partial^2+V)G=1,\quad \partial^2=\nabla^2-\partial_t^2,\label{gfe}
\ee
in matrix notation, the bodies being described by some potential
$V$.  The quantum vacuum energy is given in general by
\be
E=-\frac12\int_{-\infty}^\infty \frac{d\zeta}{2\pi} \mbox{Tr}\,\ln G G_0^{-1},
\ee
where we have subtracted out the unobservable energy of the vacuum without
material bodies, the corresponding Green's function $G_0$ being obtained from
solving Eq.~(\ref{gfe}) with $V=0$.  Here, we have introduced the 
imaginary frequency
$\zeta$, and the Green's functions are the corresponding Fourier transforms,
now satisfying more explicitly
\be
(-\nabla^2+\zeta^2+V(\mathbf{r};\zeta))G(\mathbf{r,r'};\zeta)
=\delta(\mathbf{r-r'}).
\ee
From this it is easily shown (see, for example, 
Ref.~\cite{Kenneth:2007jk,Milton:2007wz}) that
the interaction between two bodies, 1 and 2, is given by the famous 
$TGTG$ formula,
\be
E_{12}=\frac12\int\frac{d\zeta}{2\pi}\mbox{Tr}\,\ln(1-G_0T_1G_0T_2),
\label{tgtg}
\ee
where the scattering matrices for each body are
\be
T_i=V_i(1+G_0 V_i)^{-1},\quad i=1,2.
\ee
(These equations are easily extendable to the electromagnetic situation.)

In general it is nontrivial to find the scattering matrix for a body not
possessing a great deal of symmetry, such as an infinite plane, a sphere, or a
cylinder.  In the case of weak coupling, however, where the potential is
regarded as small, so that we  replace $T$ by $V$, and keep only the first
term in the expansion of the logarithm in Eq.~(\ref{tgtg}), many exact results
can be found.  Then the Casimir energy takes the simple form 
\cite{Wagner:2008qq}
\be
E_{12}=-\frac1{64\pi^3}\int (d\mathbf{r})\int (d\mathbf{r'})
\frac{V_1(\mathbf{r})
V_2(\mathbf{r'})}{|\mathbf{r-r'}|^3}.\label{3dwc}
\ee
This formula is the analog of the pairwise
summation of van der Waals or Casimir-Polder
energies in electromagnetism---see Ref.~\cite{Milton:2008vr}.
In Ref.~\cite{Wagner:2008qq} we derived several interesting examples for forces
between finite and infinite plates, including edge effects.  Here we give some
further examples involving torques.

\section{Torque on a rectangular plate} \label{sec:rec}
\begin{figure}
 \begin{center}
  \includegraphics{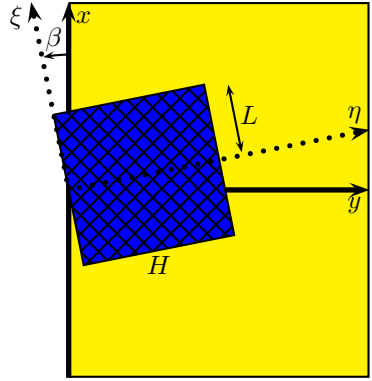}
 \end{center}
\caption{\label{fig:rot-pl}
A finite rectangular plate, of sides $H$ and $2L$, a distance $a$
above and parallel to a semi-infinite plate.
The finite plate is rotated through an angle $\beta$ about an axis 
perpendicular to both plates which passes  through the edge of both.
When $\beta=0$ the two plates are aligned, with the $2L$ side of
the upper plate directly above the edge of 
 the lower plate.  The coordinate axes
belonging to both plates are also shown: $x$ is the coordinate along the
edge of the semi-infinite plate, $y$ is the coordinate in the plate
perpendicular to the edge.  Likewise, $\xi$ is the coordinate along
the $2L$ side of the finite plate, while $\eta$ is the coordinate
in that plate perpendicular to that side.}
\end{figure}

As a first example, consider a finite, rectangular plate above a semi-infinite
plate, as shown in Fig.~\ref{fig:rot-pl}.
We assume the two plates are described by the ``semitransparent'' potentials
\begin{subequations}
\bea
V_1(\mathbf{r})&=&\lambda_1\delta(z)\theta(y),\\
V_2(\mathbf{r'})&=&\lambda_2\delta(z'-a)\theta(\eta)
\theta(H-\eta)\theta(\xi+L)\theta(L-\xi).
\eea
\end{subequations}
Here the Heaviside step function is
\be
\theta(x)=\left\{\begin{array}{cc}
                  1,&x>0,\\
0,&x<0.
                 \end{array}\right.
\ee
Thus, the first plate is a half-plane at $z=0$, described by local Cartesian
coordinates $x$ and $y$ as shown in the figure, while the second plate is 
a finite
rectangle in the $z=a$ plane, described by local Cartesian coordinates 
$\xi\in[-L,L]$ and $\eta\in[0,H]$.  
Note here that the coupling strengths on the
two plates, $\lambda_{1,2}$, have dimension of (length)$^{-1}$.  
Because the finite
plate is assumed to be rotated relative to the semi-infinite one by an angle 
$\beta$
about the $z$ axis passing through 
and perpendicular to the edges of both plates, the relation 
between the Cartesian coordinates in the two systems is
\be
x'=\xi\cos\beta+\eta\sin\beta,\quad y'=
\eta \cos\beta-\xi \sin\beta.\label{rotcoord}
\ee

We now insert these potentials into the weak-coupling formula (\ref{3dwc}), 
which gives
\be
E_{12}=-\frac{\lambda_1\lambda_2}{64\pi^3}\int_{-\infty}^\infty dx
\int_0^\infty dy
\int_{-L}^L d\xi\int_0^H d\eta\,[a^2+(x-x')^2+(y-y')^2]^{-3/2},
\ee
where the relation between $(x',y')$ and $(\xi,\eta)$ is 
given by Eq.~(\ref{rotcoord}).
We immediately carry out the integrals over the semi-infinite plate, 
with the result\footnote{This is a generic result, provided the integral
is extended over the body.  Using it, similar results can be found, for
example, for an equilateral triangle, where a cusp for $A\gg a^2$ 
appears, as expected, at $\beta=\pi/2$.  See Fig.~\ref{comp}, below.}
\be
E_{12}=-K\left[\frac{A}2+\frac1\pi
\int_{-L}^Ld\xi\int_0^H d\eta
\arctan\left(\frac\eta{a}\cos\beta-\frac\xi{a}\sin\beta\right)\right],
\label{arctanform}
\ee
in terms of the abbreviations for the area of the finite plate
 and for the magnitude of the
Casimir energy for parallel plates,
\be
A=2LH,\quad
K=\frac{\lambda_1\lambda_2}{32\pi^2a}.
\ee
The remaining integrals are straightforward, and we obtain the following exact 
result:
\be
E_{12}=-K\left(LH+\frac{a^2}{2\pi\cos\beta}
f(l,h,\beta)\right),
\ee
where, with $l=L/a$, $h=H/a$, 
\bea
f(l,h,\beta)&=&2\sin\beta (\csc^2\beta-l^2)\arctan(l\sin\beta)+2l\ln(1+l^2
\sin^2\beta)\nonumber\\
&&\quad\mbox{}+\csc\beta\left[(h\cos\beta+l\sin\beta)^2-1\right]
\arctan(h\cos\beta+l\sin\beta)\nonumber\\
&&\quad\mbox{}-\csc\beta \left[(h\cos\beta-l\sin\beta)^2-1\right]
\arctan(h\cos\beta-l\sin\beta)\nonumber\\
&&\quad\mbox{}-h\cot \beta \ln\left[\frac{1+(h\cos\beta+l\sin\beta)^2}
{1+(h\cos\beta-l\sin\beta)^2}
\right]\nonumber\\
&&\quad\mbox{}-l \ln\left[(1+h^2\cos^2\beta+l^2\sin^2\beta)^2-h^2l^2\sin^2 
2\beta\right].
\eea
When the plates are aligned, $\beta=0$, the energy reduces to
\be
E_{12}(\alpha=0)=-K\left[LH
+\frac{2LH}\pi\arctan h-\frac{aL}{\pi}\ln(1+h^2)\right]
\sim -K \left[2LH
-\frac{2La}\pi(\ln h+1)\right],
\ee 
where the last approximation holds when $H,L \gg a$, that is, the plates 
are large compared to their separation.
When the $x$ and $\xi$ axes of the plates are perpendicular, 
\be E_{12}(\beta=\pi/2)=-K LH,
\ee
with no correction, and when they are anti-aligned, $\beta=\pi$, 
so there is no overlap between the plates, 
\be
E_{12}(\beta=\pi)=-K\left[LH
-\frac{2LH}\pi\arctan h+\frac{aL}{\pi}\ln(1+h^2)\right]
\sim -K\frac{2La}\pi(\ln h+1).
\ee
The energy, which is negative, monotonically increases with the angle 
$\beta\in[0,\pi]$.

\begin{figure}
 \begin{center}
  \includegraphics[scale=.9]{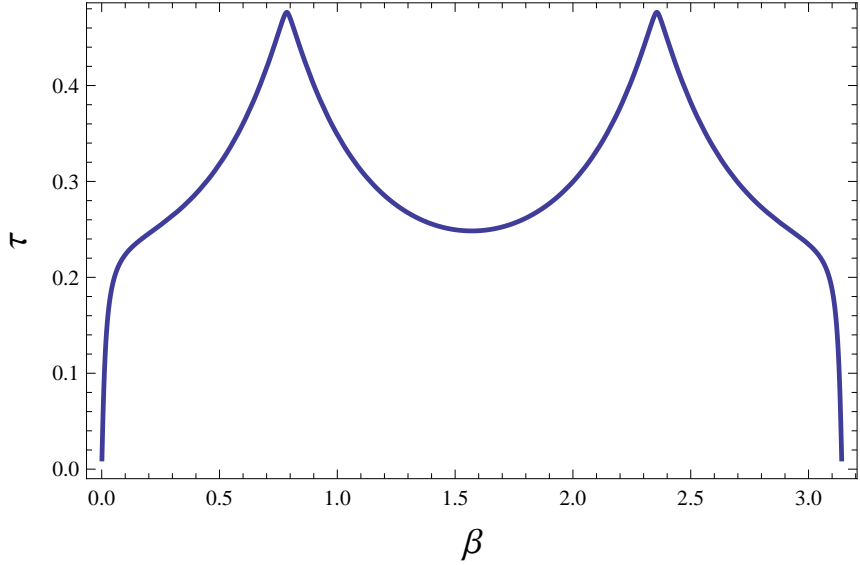}
 \end{center}
\caption{\label{fig:rec-cusp}
Torque exerted between the horizontally movable plates shown in 
Fig.~\ref{fig:rot-pl}. The torque is plotted in units of
 $-K A$, $A$ being the area of the finite plate,
$A=2HL$. The graph is for $L=100a$, $H=100a$.}
\end{figure}

It is more interesting to look at the torque,
rather than plot the energy, 
\be
\tau=-\frac\partial{\partial\beta}E_{12}(\beta).\label{torqueformula}
\ee
Figure \ref{fig:rec-cusp} shows a typical case.
Even though the energy is a rather smooth function of $\beta$, 
oscillating around a straight line whose slope is the negative of
the average torque, 
the torque exhibits prominent approximate cusps.  
The other striking feature of the plot is its symmetry about $\beta=\pi/2$. 
This reflects the antisymmetry of the integral in the
 starting point for the energy, Eq.~(\ref{arctanform}), under
$y'\to-y'$, which implies that $\tau(\beta)=\tau(\pi-\beta)$.

\begin{figure}
 \begin{center}
  \includegraphics[scale=.9]{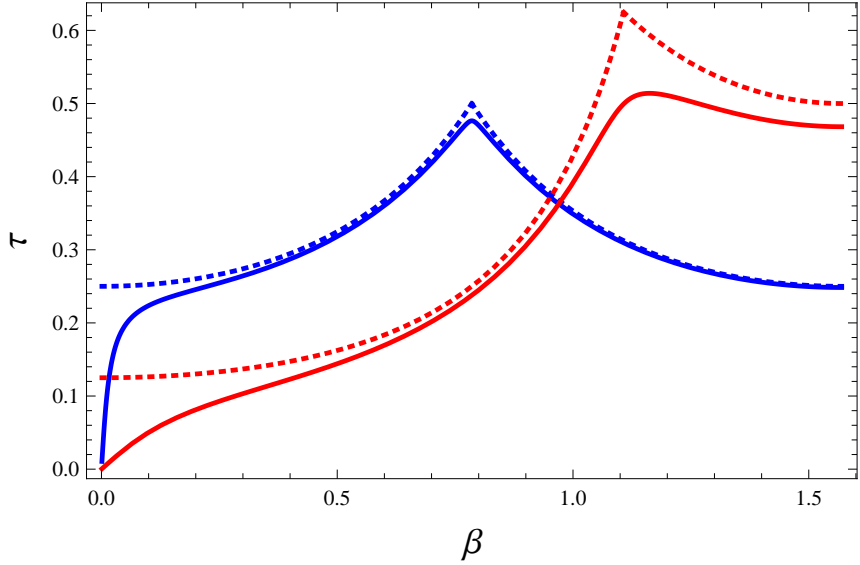}
 \end{center}
\caption{\label{t100100} Exact torque (solid curves) compared with the 
PFA torque (dotted curves). Again the torque is given in units of $-K A$. 
The first set of curves (blue) with cusp at $\pi/4$ is for
a rectangle  $L=100 a$, 
$H=100 a$, while the second set (red) with the cusp at $\beta=\arctan 2 =1.107$
 is for a square with  $L=10 a$, $H=20  a$.}
\end{figure}

There is an easy way to understand the structure seen in 
Fig.~\ref{fig:rec-cusp}, which is almost entirely geometrical.  That is the
proximity force approximation (PFA) \cite{pfa}, 
which says here that only overlapping plate elements
contribute, and that then one should use for the energy per unit area the 
``Casimir'' energy per unit area for infinite, parallel plates, here
\be
E_C=-\frac{\lambda_1\lambda_2 }{32\pi^2 a}=-K.
\ee
The PFA  energy is then 
\be
E_{{\rm PFA}}=E_C A_o,
\ee
where $A_o$ is the area of overlap.
Here, the overlap area depends on which region of $\beta$ one is in:
\begin{subequations}
\bea
\beta\in[0,\arctan H/L]:\quad  A_o&=&2HL-\frac12 L^2\tan\beta,\\
\beta\in[\arctan H/L,\pi-\arctan H/L]: \quad A_o&=&HL+\frac12 H^2\cot\beta,\\
\beta\in[\pi-\arctan H/L,\pi]:\quad A_o&=&-\frac12 L^2 \tan\beta.
\eea
\end{subequations}
We compare the exact result for the torque with the PFA torque,
\be
\tau_{\rm PFA}=-\frac{K}{2}\left\{
\begin{array}{cc}
L^2\sec^2\beta,&\beta\in[0,\arctan H/L] \,\,\mbox{or}\,\,
\beta\in[\pi-\arctan H/L,\pi],\\ 
H^2\csc^2\beta,&\beta\in[\arctan H/L,\pi-\arctan H/L],
\end{array}\right.
\ee
in Fig.~\ref{t100100}.

It is seen that there is very little difference between the exact torque and
the PFA for moderate values of $H/a$, $L/a$, except for $\beta$ very close to
0 (or $\pi$), where the exact torque vanishes. For very small angles 
(mod $\pi$) the large length approximation breaks down due to the 
multiplication by a small sine function.
 Evidently, the cusps arise
when the corners of the finite rectangle pass over the edge of the 
semi-infinite plate.  The PFA torque does not vanish at $\beta=0$ because
the area of overlap varies linearly with $\beta$ for small $\beta$.

\section{Torque on a  disk}
\begin{figure}
\begin{center}
\includegraphics{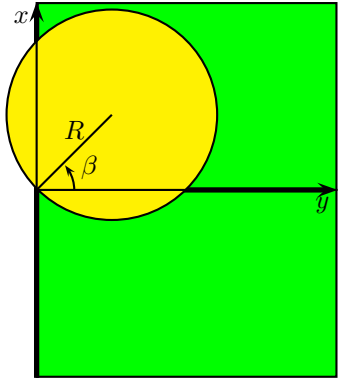}
\end{center}
\caption{\label{disk} A  disk of radius $R$ a distance $a$ above a 
semi-infinite plate.  Both objects are described by semitransparent 
$\delta$-function potentials.
The diameter of the disk makes an angle $\beta$
with respect to the planar
normal to the edge of the disk, and it is imagined
that the disk is free to rotate about an axis perpendicular to the plane of
both objects and passing through the edge of both.}
\end{figure}

To contrast with the above calculation, we consider a semitransparent disk
of radius $R$ a distance $a$ above a semitransparent plate, as illustrated
in Fig.~\ref{disk}.
We consider the disk as free to rotate about an axis passing through the
boundary of both objects.  
The angle of rotation $\beta$ is  so chosen that $\beta=0$
corresponds to the disk entirely lying above the semi-infinite plate, which is
the equilibrium position, so that for $0<\beta<\pi$ a negative Casimir
torque tends to reduce the angle.  

\begin{figure}
\begin{center}
\includegraphics[scale=.9]{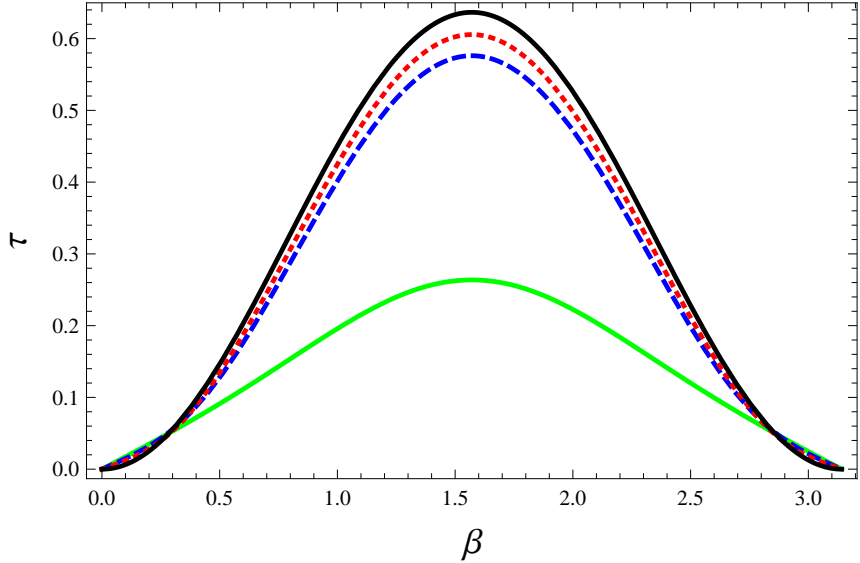}
\end{center}
\caption{\label{disk-torque}  Torque between a disk above a half-plate,
as shown in Fig.~\ref{disk}. The torque, in units of $-K A$, $A=\pi R^2$
being the area of the disk,
is plotted as a function of the angle $\beta$.
The lower (green) curve shows the
torque for $R=a$, the second (dashed blue) curve shows the torque for
$R=10a$, the third (dotted red) curve shows the torque for $R=20a$,
and the top (solid black) curve shows the proximity force
approximation.}
\end{figure}

In weak coupling, the energy is given by
the analog of Eq.~(\ref{arctanform}),
\be
E_{12}=-K\left[\frac{\pi R^2}2+\frac1\pi\int_0^R d\rho\,\rho\int_0^{2\pi}
d\phi\arctan\frac{y'}a\right],
\ee
where in terms of polar coordinates with origin at the center of the disk,
\be
y'=R\cos\beta+\rho\cos\phi.
\ee
The integrals are straightforward (although Mathematica \cite{wolfram}
 has trouble dealing
with the branches), and the following is the result for the weak-coupling
energy for this configuration:
\bea
E&=&-\frac{K\pi R^2}2\bigg\{1+\frac1{\pi i}\ln\left(
\frac{1+i r\cos\beta+\sqrt{1+2 i r\cos\beta+r^2\sin^2\beta}}
{1-i r\cos\beta+\sqrt{1-2 i r\cos\beta+r^2\sin^2\beta}}\right)\nonumber\\
&&\quad\mbox{}-\frac4{\pi r}\cos\beta+\frac1{\pi ir^2}\left[
\sqrt{1+2 i r\cos\beta+r^2\sin^2\beta}-
\sqrt{1-2 i r\cos\beta+r^2\sin^2\beta}\right]\nonumber\\
&&\quad\mbox{}+\frac{\cos\beta}{\pi r}\left[
\sqrt{1+2 i r\cos\beta+r^2\sin^2\beta}
+\sqrt{1-2 i r\cos\beta+r^2\sin^2\beta}\right]\bigg\},
\eea
with $r=R/a$. As $r\to\infty$, $0<\beta<\pi$, this tends to the PFA
energy,
\be
E_{\rm PFA}=-K\frac{\pi R^2}2\left(2-\frac{2\beta}\pi+\frac{\sin2\beta}
\pi\right),
\ee
where the coefficient of $-K$ is simply the area
of overlap of the disk above the
lower plate.  The corresponding torque is
\be
\tau_{\rm PFA}=-2K R^2\sin^2\beta,\quad \beta\in[0,\pi].
\ee
The exact torque is compared with the PFA in Fig.~\ref{disk-torque}.
Note that there are no cusps here because of the absence of sharp
corners in the disk.  Both the exact torque and the PFA vanish at 
$\beta=0$, $\pi$, but the PFA torque (unlike the exact torque) has zero 
slope there because the overlap energy varies like $\beta^3$ for small $\beta$.
We see, once again,
 that the PFA is very accurate as long as the size of the disk is large
compared to the separation, $R\gg a$. 

\begin{figure}
\includegraphics[scale=.9]{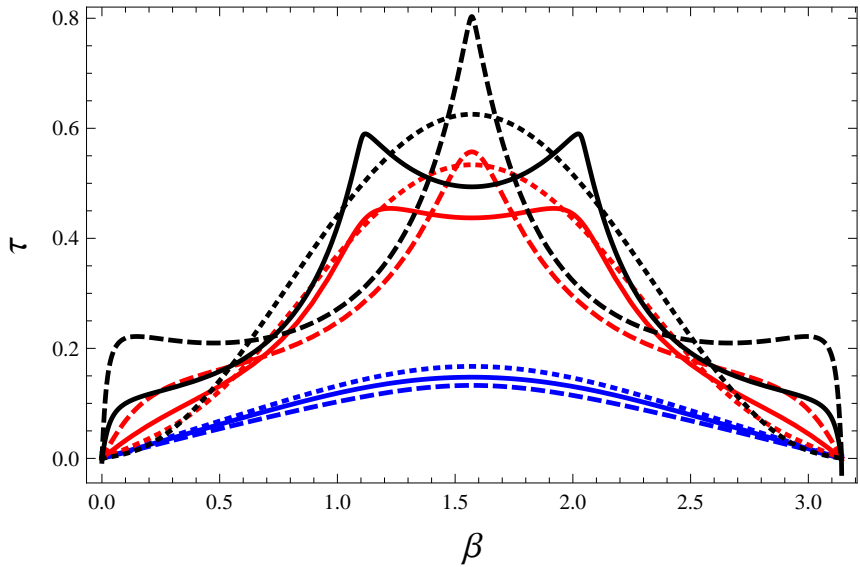}
\caption{\label{comp} The torque on a disk,  a square, and an
equilateral triangle  of the same area $A$
as a function of the angle $\beta$.  The torque is plotted in units of $-KA$.
The solid curves are for the square, the long-dashed curves 
for the equilateral
triangle, and the short-dashed curves for the disk.  The lower set
(blue) is for $A=a^2$, the middle set (red) is for $A=100 a^2$, and the top
set (black) is for $A=10^4 a^2$.}
\end{figure}

We compare the torque on a disk, a square, and an equilateral triangle 
(expressions not given here) of equal areas in 
Fig.~\ref{comp}.
It is seen that as the area increases, the distinction between the shapes
 grows more pronounced, with cusps appearing for a square at $\beta=\arctan 2$
and $\beta=\pi-\arctan 2$, and for a triangle at $\beta=\pi/2$, both having
a distinct shoulder near $\beta=0,\pi$.  The average torque is much the
same. The lower set of curves in Fig.~\ref{comp} seems to suggest that for
sufficiently small areas the distinction between the shapes of the upper
body becomes irrelevant; 
in particular, it is noteworthy that for  $A<21.3a^2$,
the cusps for the square  entirely disappear. And although there is always
a maximum in the torque for $\beta=\pi/2$ for the triangle, the cusp-like
character disappears for $A<2a^2$. 
\section{Annular piston}

As a third torque example, we are inspired by our recent work 
\cite{Milton:2013yqa,Milton:2013xia} 
involving
an annular piston, in which two radial plates between concentric cylinders
are free to move under the influence of quantum vacuum forces contained
within the annular sector so defined. Those investigations, in turn,
were inspired by a suggestion that the relation between torque
and energy might not be the expected one (\ref{torqueformula})
when divergent self-energies
are involved  \cite{Fulling:2012wa}.  Although we showed, in fact,
that the renormalized energy does not suffer that defect
\cite{Milton:2013yqa,Milton:2013xia}, the issue is utterly irrelevant
here because the interaction energy of distinct bodies is completely finite.

\begin{figure}
 \begin{center}
  \includegraphics{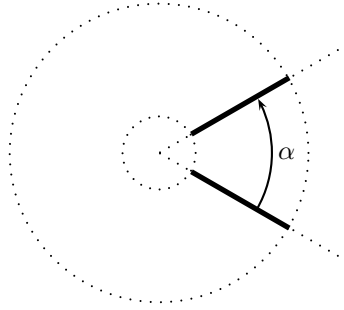}
 \end{center}
\caption{\label{annpist} Two immaterial concentric cylinders are intersected 
with radial plates (ribbons).  
The Casimir torque between the two radial plates,
separated by an angle $\alpha$, is to be calculated.  The inner radius is $a$,
the outer radius $b$.}
\end{figure}

\begin{figure}
 \begin{center}
  \includegraphics[scale=.9]{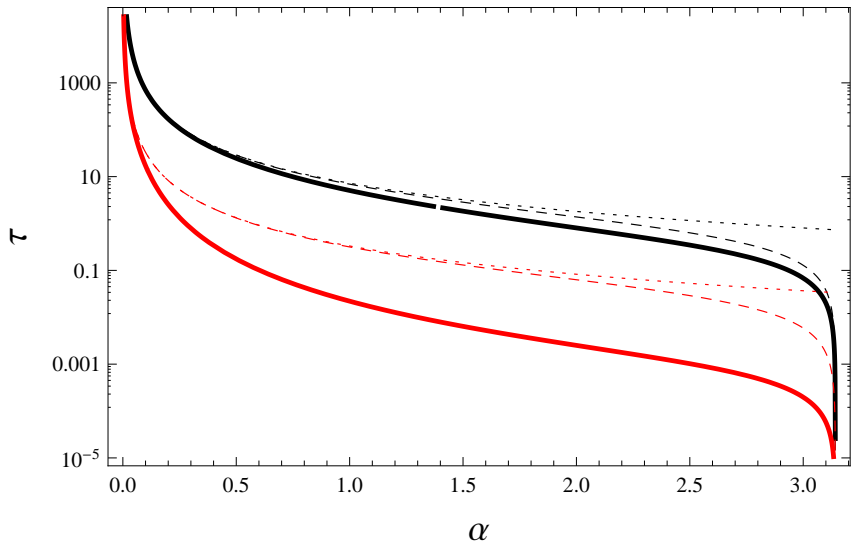}
 \end{center}
\caption{\label{fig:aptorque} 
The Casimir torque (in units of $-\mathcal{K}/\pi$) 
in the piston geometry shown in Fig.~\ref{annpist}.
This is computed from Eq.~(\ref{anntorqueenergy}) (solid curves)
 and compared to the PFA
torque (\ref{pfa:annpist})
(dotted curves).  The upper set of curves (black) is for $a/b=0.1$, 
the lower set (red)
for $a/b=0.9$.  In general, the PFA is rather similar to the true value,
and approaches it for small $\alpha$.  The improved version of the 
PFA (\ref{improvedpfa})
(dashed curves) more closely matches the true torque for large angles.
}
\end{figure}

Here we abstract from that annular sector calculation
 and consider only the plates, free to slide
on the circular cylindrical tracks as shown in Fig.~\ref{annpist}.
Here the potentials of the two plates are taken to be, in cylindrical 
coordinates,
\begin{subequations}
\bea
V_1(\mathbf{r})&=&\lambda_1 \frac1\rho\delta(\phi)\theta(\rho-a)\theta(b-\rho),\\
V_2(\mathbf{r'})&=&\lambda_2 \frac1{\rho'}\delta(\phi'-\alpha)\theta(\rho'-a)
\theta(b-\rho').
\eea
\end{subequations}
The $\rho^{-1}$ factors, required for dimensional consistency,
are necessary for keeping the weighting of elements along the plates
constant, as perhaps most easily seen by doing the calculation in two
rotated Cartesian coordinate systems as in Sec.~\ref{sec:rec}.
When these potentials
are inserted into Eq.~(\ref{3dwc}), and the trivial integrals over
$z$ (the direction of the axis of the cylinders) and $\phi$ and $\phi'$ carried
out, we have for the energy per unit length
\be
\mathcal{E}=-\frac{\mathcal{K}}{\pi}
\int_a^b d\rho\int_a^b d\rho'\frac1{\rho^2+\rho^{\prime2}
-2\rho\rho'\cos\alpha},\quad \mathcal{K}=\frac{\lambda_1\lambda_2}{32\pi^2}.
\ee
Such integrals were carried out in Ref.~\cite{Wagner:2008qq}, and indeed the 
general result for two ribbons at an angle is given there in terms of 
inverse tangent
integral functions, but no plots are shown.  For this situation, the energy
per unit length can be written as
\be
\mathcal{E}=-\frac{\mathcal{K}}\pi \csc\alpha\left[
G(a/b,\alpha)-\pi \ln\left(\frac{a}b\right) \theta\left(\arccos\frac{2a/b}
{(a/b)^2+1}-\alpha\right)\right].\label{annpisten}
\ee
Here, in terms of principal-value functions as defined internally in 
Mathematica \cite{wolfram},
\bea
G(a/b,\alpha)&=&-\frac\alpha2\ln\left[
\frac{(2ab-(a^2+b^2)\cos\alpha)^2+(a^2-b^2)^2
\sin^2\alpha}{4a^2b^2(1-\cos\alpha)^2}\right]\nonumber\\
&&\quad\mbox{}-\ln\frac{a}b \arctan\left(
\frac{(b^2-a^2)\sin\alpha}{2ab-(a^2+b^2)\cos\alpha)}\right)\nonumber\\
&&\quad\mbox{}+\frac1{2i}\bigg[
\mbox{Li}_2\left(1-\frac{a}b e^{-i\alpha}\right)
-\mbox{Li}_2\left(1-\frac{a}b e^{i\alpha}\right)
+\mbox{Li}_2\left(1-\frac{b}a e^{-i\alpha}\right)
-\mbox{Li}_2\left(1-\frac{b}a e^{i\alpha}\right)\nonumber\\
&&\quad\mbox{}+2\mbox{Li}_2\left(1- e^{i\alpha}\right)
-2\mbox{Li}_2\left(1- e^{-i\alpha}\right)\bigg],\label{anntorqueenergy}
\eea
in terms of the dilogarithm function $\mbox{Li}_2(z)$ \cite{lewin,lewin2}. 
The step-function term in Eq.~(\ref{annpisten}) is inserted so that the 
function is continuous, and has the correct behavior as $\alpha\to 0$.
Again from this we calculate the torque, which is shown in 
Fig.~\ref{fig:aptorque}.
The graph shows that the exact torque is rather similar to that obtained
by the PFA, which becomes accurate only at small angles $\alpha$.
The PFA energy is, in the small angle form, 
\be
\mathcal{E}_{\rm PFA}=-\mathcal{K}\int_a^b
\frac{d\rho}{\rho\alpha}
=-\mathcal{K}\frac1\alpha\ln\frac{b}a,
\ee
so the corresponding torque is
\be
\tau_{\rm PFA}=-\mathcal{K}\frac1{\alpha^2}\ln\frac{b}a.
\label{pfa:annpist}
\ee
An improved version of the PFA is based on bisecting the angle $\alpha$,
and connecting elements on the two plates equally above and below
the bisector.  This gives
\begin{subequations}
\bea
\mathcal{E}_{\rm PFA}'&=&-\mathcal{K}\frac1{2\sin\alpha/2}\ln\frac{b}a,\\
\tau_{\rm PFA}'&=& -\frac{\mathcal{K}}4\frac{\cos\alpha/2}{\sin^2\alpha/2}\ln\frac{b}a.
\label{improvedpfa}
\eea
\end{subequations} 
This approximation is also shown in Fig.~\ref{fig:aptorque}; for large angles, it
is considerably more accurate than the version in Eq.~(\ref{pfa:annpist}), and
correctly vanishes at $\alpha=\pi$.  The approximation becomes better for
smaller values of $a/b$, because the separation is then smaller compared
to the width of the ribbons.

\section{Discussion}
In this paper we have discussed some examples of Casimir torque,
which can be dealt with entirely analytically because we are working
in the weak-coupling approximation.  For simplicity, we have considered
a massless scalar field, interacting with $\delta$-function potentials.
We have examined planar objects, specifically a rectangle above
and parallel to  a semi-infinite
plate, rotated relative to each other about a common normal to the plates,
an equilateral triangle and a disk above a half-plate,
and two ribbons inclined relatively to each other in an 
annular piston geometry.
An interesting observation is that when the size of the finite object
is comparable to the separation distance, the shape dependence starts becoming
irrelevant.  In the opposite limit, when the size of the object is large compared
to the separation, the exact results in each case rather closely match the approximate torques
found using the proximity force approximation.  This gives us some confidence
in the use of the latter for more realistic situations, involving conductors
and dielectrics interacting through the quantum electrodynamic vacuum, where
exact calculations of energies and torques are difficult to obtain.
Unfortunately, corrections to the PFA are known only in the case of smooth
deformations \cite{Fosco:2011xx}, 
not for objects with sharp edges, such as considered here.
Of course, there are a number of recent discussions of edge effects
in strong coupling (Dirichlet or perfect conducting boundaries),
for example, for a half-plane above an infinite one
\cite{Karabali:2013qza,Gies:2006xe}, and for various 
sharp-edged objects \cite{Graham:2011ta, Maghrebi:2010wp}, 
including discussion of torques between such
objects \cite{Emig:2008qr, Graham:2009zb}.  These previous
discussions were largely numerical, so our weak-coupling
analytic calculations provide valuable insight.



\begin{thebibliography}{99}
%
%
%
\bibitem{Casimir:1948dh} 
  H.~B.~G.~Casimir,
{\it Kon.\ Ned.\ Akad.\ Wetensch.\ Proc.}\ {\bf 51}, 793 (1948).

%
\bibitem{Lifshitz:1956zz} 
  E.~M.~Lifshitz,
{\it Zh.\ Eksp.\ Teor.\ Fiz.}\ {\bf 29}, 94 (1956)
 [{\it Sov.\ Phys.\ JETP\/} {\bf 2}, 73 (1956)].


\bibitem{dzyaloshinskii}
I. D. Dzyaloshinskii, E. M. Lifshitz, and L. P. Pitaevskii, 
{\it Usp.\ Fiz.\ Nauk}, {\bf 73}, 381 (1961) [{\it Sov.\ Phys.\ Usp.}\ {\bf 4},
 153 (1961)].


%
\bibitem{Bordag:2009zz} 
  M.~Bordag, G.~L.~Klimchitskaya, U.~Mohideen, and V.~M.~Mostepanenko,
 {\it Advances in the Casimir Effect,}
 International series of monographs on physics, No.~145, 
Oxford University Press, 2009.

\bibitem{dalvitbook}
D. Dalvit, P.  Milonni, D. Roberts, and F.  da Rosa,  (Eds.)
{\it Casimir Physics}, Lecture Notes in Physics, Vol. 834,
Springer, Berlin, 2011.

\bibitem{Lombardo:2008zza} 
  F.~C.~Lombardo, F.~D.~Mazzitelli, and P.~I.~Villar,
{\it  J.\ Phys.\ A} {\bf 41}, 164009 (2008).

\bibitem{CaveroPelaez:2008tk} 
  I.~Cavero-Pelaez, K.~A.~Milton, P.~Parashar, and K.~V.~Shajesh,
{\it  Phys.\ Rev.\ D }{\bf 78}, 065019 (2008)
  [arXiv:0805.2777 [hep-th]].

\bibitem{barash} Y. Barash, {\it Izvestiya V. U. Z., Radiofizika\/} {\bf 16}, 
1227 (1973)  [{\it Sov.\ Radiophys.} {\bf 16}, 945 (1973).

\bibitem{barash2}
D. Iannuzzi, J. N. Munday, Y. Barash, and F. Capasso
{\it Phys.\ Rev.\ A} {\bf71}, 042102 (2005)
[arXiv:quant-ph/0410136].


\bibitem{lateral1}
H.-C. Chiu, G.L. Klimchitskaya, V.N. Marachevsky, V.M. Mostepanenko, 
and U. Mohideen, {\it Phys.\ Rev.\ B} {\bf80}, 121402(R) (2009)
 [arXiv:0909.2161].


\bibitem{lateral2}
H.-C. Chiu, G. L. Klimchitskaya, V. N. Marachevsky, V. M. Mostepanenko, 
and U. Mohideen,
{\it Phys.\ Rev.\ B}  {\bf81}, 115417 (2010)
[arXiv:1002.3936].

\bibitem{Kenneth:2007jk} 
  O.~Kenneth and I.~Klich,
 {\it Phys.\ Rev.\ B\/ } {\bf 78}, 014103 (2008)
  [arXiv:0707.4017 [quant-ph]].

\bibitem{Milton:2007wz} 
  K.~A.~Milton and J.~Wagner,
 {\it J.\ Phys.\ A }{\bf 41}, 155402 (2008)
  [arXiv:0712.3811 [hep-th]].

\bibitem{Wagner:2008qq} 
  J.~Wagner, K.~A.~Milton, and P.~Parashar,
 {\it  J.\ Phys.\ Conf.\ Ser.} {\bf 161}, 012022 (2009)
  [arXiv:0811.2442 [hep-th]].

\bibitem{Milton:2008vr} 
  K.~A.~Milton, P.~Parashar, and J.~Wagner,
{\it  Phys.\ Rev.\ Lett.} {\bf 101}, 160402 (2008)
  [arXiv:0806.2880 [hep-th]].

\bibitem{pfa} B. V. Derjaguin, {\it Kolloid Z.} {\bf 69}, 155 (1934).

\bibitem{wolfram}
Wolfram Research, Inc., Mathematica, Version 8.0, Champaign, IL (2010).
%
\bibitem{Milton:2013yqa} 
  K.~A.~Milton, F.~Kheirandish, P.~Parashar, E.~K.~Abalo, S.~A.~Fulling, 
J.~D.~Bouas, H.~Carter, and K.~Kirsten,
{\it Phys.\ Rev.\ D\/} {\bf88}, 025039 (2013)  [arXiv:1306.0866 [hep-th]].

\bibitem{Milton:2013xia} 
  K.~A.~Milton, P.~Parashar, E.~K.~Abalo, F.~Kheirandish, and K.~Kirsten,
{\it Phys.\ Rev.\ D\/} {\bf 88}, 045030 (2013)
  [arXiv:1307.2535 [hep-th]].
%
\bibitem{Fulling:2012wa} 
  S.~A.~Fulling, F.~D.~Mera, and C.~S.~Trendafilova,
{\it  Phys.\ Rev.\ D\/} {\bf 87}, 047702 (2013)
  [arXiv:1212.6249 [hep-th]].

\bibitem{lewin}
L. Lewin, {\it Dilogarithms and Associated Functions},  Macdonald,
London, 1958.
\bibitem{lewin2}
L. Lewin, {\it Polylogarithms and Associated Functions}, North-Holland, 
New York, 1981.

%
\bibitem{Fosco:2011xx} 
  C.~D.~Fosco, F.~C.~Lombardo, and F.~D.~Mazzitelli,
{\it  Phys.\ Rev.\ D\/} {\bf 84}, 105031 (2011)
  [arXiv:1109.2123 [hep-th]].

\bibitem{Karabali:2013qza} 
  D.~Karabali and V.~P.~Nair,
{\it Phys.\ Rev.\ D\/} {\bf 87}, 105021 (2013)
  [arXiv:1304.0511 [hep-th]].

\bibitem{Gies:2006xe} 
  H.~Gies and K.~Klingmuller,
 {\it Phys.\ Rev.\ Lett.\ }  {\bf 97}, 220405 (2006)
  [quant-ph/0606235].

\bibitem{Graham:2011ta} 
  N.~Graham, A.~Shpunt, T.~Emig, S.~J.~Rahi, R.~L.~Jaffe, and M.~Kardar,
{\it  Phys.\ Rev.\ D} {\bf 83}, 125007 (2011)
  [arXiv:1103.5942 [quant-ph]].

\bibitem{Maghrebi:2010wp} 
  M.~F.~Maghrebi, S.~J.~Rahi, T.~Emig, N.~Graham, R.~L.~Jaffe, and M.~Kardar,
  {\it Proc.\ Nat.\ Acad.\ Sci.\/ } {\bf 108}, 6867 (2011)
  [arXiv:1010.3223 [quant-ph]].

\bibitem{Emig:2008qr} 
  T.~Emig, N.~Graham, R.~L.~Jaffe, and M.~Kardar,
{\it  Phys.\ Rev.\ A\/} {\bf 79}, 054901 (2009)
  [arXiv:0811.1597 [cond-mat.stat-mech]].

\bibitem{Graham:2009zb} 
  N.~Graham, A.~Shpunt, T.~Emig, S.~J.~Rahi, R.~L.~Jaffe, and M.~Kardar,
{\it  Phys.\ Rev.\ D\/} {\bf 81}, 061701 (2010)
  [arXiv:0910.4649 [quant-ph]].


\end{thebibliography}
\end{document}